\documentclass[doublecol]{epl2}
\usepackage{graphics}
\usepackage{subfigure}
\usepackage{diagbox}

\title{{Game among interdependent networks: the impact of rationality on system robustness}}
\shorttitle{Game among interdependent networks}

\author{Yuhang Fan\inst{1,2} \and Gongze Cao\inst{2} \and Shibo He\inst{1} \and Jiming Chen\inst{1} \and Youxian Sun\inst{1}}
\institute{
  \inst{1} State Key Lab. of Industrial Control Technology, Zhejiang University - Hangzhou, 310027, China\\
  \inst{2} School of Mathematics, Zhejiang University - Hangzhou, 310027, China
}


\abstract{
Many real-world systems are composed of interdependent networks that rely on one another. Such networks are typically designed and operated by different entities, who aim at maximizing their own {payoffs. There exists a game among these entities when designing their own networks.} In this paper, we study the game investigating how the rational behaviors of entities impact the system {robustness}. We first introduce a mathematical model to quantify the interacting payoffs among varying entities. Then we study the Nash equilibrium {of the game} and compare it with the optimal social welfare. We reveal that the cooperation among different entities can be reached to {maximize the social welfare in continuous game} only when the average degree of each network is constant. {Therefore,} the huge gap between Nash equilibrium and optimal social welfare {generally exists}. The rationality of entities {makes the system inherently deficient} and even {renders it extremely vulnerable} in some cases. {We analyze our model for two concrete systems with continuous strategy space and discrete strategy space, respectively.} Furthermore, we uncover some factors (such as weakening coupled strength of interdependent networks, designing suitable topology dependency of the system) that help reduce the gap and the system vulnerability.
}

\begin{document}

\maketitle

In interdependent networks, nodes from {different networks} rely on {one another}. The failure of a node in one network causes its dependent {nodes} to also fail, leading to an iterative cascade of {failures}. They are, consequently, much more fragile than independent networks \cite{catastrophic,Networks,kenett2015networks}. Much attention has been paid on how interdependency results in the catastrophic cascade of {failures} and how to improve {the robustness of such systems} \cite{Towards,Reduce,Multisupport,Robustness,Recovery,Recentadvances}.

{The game, especially evolutionary game, on interdependent networks has been well studied before \cite{wang2012evolution,wang2013interdependent,wang2015evolutionary}. Some methods, such as optimizing the interdependence topology \cite{wang2013optimal} and designing proper cooperative mechanisms \cite{wang2014self}, have been proposed to promote the cooperation on interdependent networks. Further relevant research has {been done under the} elevated levels of cooperation in interdependent networks {more precisely}, particularly in relation to information transfer \cite{jiang2013spreading,szolnoki2013information}.}

{However, these studies, mainly based on conventional networked evolutionary game \cite{nowak1992evolutionary,hauert2004spatial}, regarded nodes in networks as the players of the game and represented the mutilayer relation between them via interdependent networks. So far{,} little {is} known about the game among interdependent networks in which the players are networks themselves.} In practice, different networks are typically designed and operated by varying entities \cite{Economic,Modelling,Identifying}. For example, {the power and communication networks} are owned by different companies in China. Each entity aims at maximizing its own {payoff} when building the network, without considering the overall {system} performance. Clearly, there {exists} a game during the formation of the interdependent networks, where a network is taken as a player. Studying such a game helps understand how the topology of the practical interdependent networks {is} formed, and {provides} insight into their inherent performance degradation, which has not been studied yet.

We introduce a mathematical framework based on random graph theory and percolation theory \cite{Percolation} for studying this game. The system is composed of $n$ interdependent networks {$N_i,i \in\{1,2,\dots,n\}$,} and the dependency is fixed. After a fraction $1-p_i$ of nodes {being randomly} removed from network $N_i$, there is an iterative cascade of {failures}. We denote the fraction of nodes in {the} giant component {of} network $N_i$ by $P_{\infty,i}$ (when the number of nodes approaches infinity, it represents the probability of the existence of the giant component) which is a function of {$p_j, j \in\{1,2,\dots,n\}$}. Let $\langle k \rangle _i$ be the average degree of network $N_i$. The payoff of $N_i$ is the difference of the income $I_i$  and {the} cost $O_i$ associated with {building and operating} the network {$N_i$}. Clearly, the income {of $N_i$, denoted by $I_i(P_{\infty},i)$,} is positively correlated to {the ratio of functional nodes (i.e., the fraction of nodes in mutually connected giant component $P_{\infty,i}$ \cite{catastrophic} which is positively correlated to the robustness of the network) in practice}. As $P_{\infty,i}$ is a function of $p_j$, we can regard $I_i$ as a function of $p_j$. Since {it requires extra cost to construct and operate links}, we assume that the cost function $O_i$ increases with the average degree $\langle k \rangle_i$.

\begin{figure}[h]\label{maxY}
  \centering
  \includegraphics[scale=0.25,width=6cm]{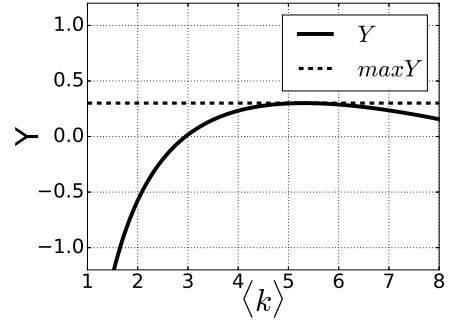}
  \caption{{(Colour on-line)} $Y$, the social welfare of the system composed from two interdependent ER networks with both average degree being $\langle k \rangle$ is shown as a function of $\langle k \rangle$ {for cost function $O(\langle k \rangle)=0.08\langle k \rangle$}. Here, {for simplicity,} the distribution of $p$, i.e., the {fraction of remaining nodes after the attack,} is supposed to {have a} uniform distribution. {It can be seen from the figure that the social welfare is not a monotone function of the average degree and gets its optimum at $\langle k \rangle=5.54$}.}\label{figure2}
\end{figure}

{Because} the fraction $p_i$ of nodes that {remain} in {the} network $N_i$ {after the initial attack} is not constant in {the} real world and {is} affected by many factors (such as weather condition for power network \cite{Modellinginterdependent}), we suppose it {has} a probability distribution {with density} $\psi_i$, that is, $P(p_i<\alpha)=\int_0^\alpha \psi_i(x)\mathrm{d}x$. {With this,} we can compute the mathematical expectation $E(I_i)$ of the income $I_i$. The payoff $Y_i$ of the network $N_i$ is thus:
\begin{equation}
Y_i=E(I_i(P_{\infty,i}))-O_i(\langle k \rangle _i) \label{eq:payoff_i},
\end{equation}
{and} the payoff $Y$ of the whole system (a.k.a. social welfare) is
\begin{equation}
Y=\sum_{i=i}^n Y_i \label{eq:payoff_system}.
\end{equation}
Note that increasing $\langle k \rangle$ may not always improve {the} social welfare since both the income and the cost increase with $\langle k \rangle$. In a cooperative system, an optimal $\langle k \rangle$ and {the corresponding strategy} can be calculated to maximize the social welfare. {To be more comprehensible, we} {provide} an example of two {totally} interdependent ER networks with the same average degree $\langle k \rangle$ \cite{Random} {and a fraction of $1-p_i$ nodes randomly {being} removed from network $N_i$. According to the one-to-one correspondence \cite{onecorrespondence}, it is equivalent to removing $1-p$ {(which equals to $1-p_1p_2$)} fraction of nodes from one network. There is a first order percolation transition with the threshold $p_c$.} It has been shown that the threshold $p_c=2.4554/\langle k \rangle$ \cite{catastrophic}. We assume, for simplicity, that the income {$I=0$} when $p<p_c$, and $I=1$ when $p>p_c$, and that {$p$}, the fraction of {functional} nodes {in $N_1$ after the initial attack,} follows a uniform distribution. The cost function $O$ is set to be a linear function with coefficient $0.08$. Then, the social welfare $Y=(1-p_c)-0.08\langle k \rangle =1-(2.4554/\langle k \rangle+0.08\langle k \rangle)$. Clearly, {a} higher $\langle k \rangle$ leads to {a} higher income as well as {a} higher cost. It is easy to compute the {social optimum} at $\langle k \rangle=5.54$ {where the optimal social welfare} $Y=0.11$ \cite{Gametheory,Aprimer}.(see FIG.\ref{figure2})

Unfortunately, the average degree is {typically} implicit in $p_c$ and $P_{\infty,i}$, which makes {our} analysis complicated. {Here,} we introduce real variables $\{q_i\},{i\in\{1,\dots,n\}},$ as strategy profiles to quantify the construction pattern of network $N_i$ by the entity $i$. We suppose that the strategy set of entity $i$ is $Q_i$, i.e., the range of variable $q_i$, and $Q=Q_{1}\times\cdots\times Q_{n}$ is the set of strategy profiles in this game. Let $q_{-i}$ be the strategy profile of all entities except for player $i$. According to \cite{Fractal}, the degree distribution of a node in network $N_i$ can be characterized by $P_i(k,q_i)$. For instance, $q_i$ could be {the} average degree {of an} ER network decided by entity $i$. We introduce the generating function of network $N_i$ whose arguments are $x$ and $q_i$ as
\begin{equation}
G_i(x,q_i)=\sum_{k=0}^{\infty} P_i(k,q_i)x^k. \label{eq:Gi}
\end{equation}
Different from previous studies \cite{Fractal}\cite{Spread}, the strategy {parameters} $q_i$ {are} included in the generating function to assist the analysis of the game. The average degree $\langle k\rangle_i$ of network $N_i$ can be calculated as
\begin{equation}
\langle k\rangle_i=\frac{\partial}{\partial x}G_i(1,q_i)=\sum_{k=0}^{\infty} kP_i(k,q_i), \label{eq:ki}
\end{equation}
which is a continuous function of $q_i$. Analogously, we introduce the generating function of the underlying branching process {\cite{Networks}}
\begin{equation}
H_i(x,q_i)=\frac{G'_i(x,q_i)}{G'_i(1,q_i)}, \label{eq:Hi}
\end{equation}
where $G'_i(x,q_i)$ is the derivative of $G$ with respect to $x$. The probability that a randomly chosen surviving {node} belongs to {the} giant component {\cite{catastrophic,Robustness}} is given by
\begin{equation}
  g_i(p,q_i)=1-G_i[pf_i(p,q_i)+1-p,q_i], \label{eq:gi}
\end{equation}
where $f_i$ satisfies
\begin{equation}
  f_i(p,q_i)=H_i[pf_i(p,q_i)+1-p,q_i] \label{eq:fi}.
\end{equation}

Similar to {Kirchhoff equations}, for fixed $q_i$, we can {arrive at a system of iterative equations of} unknowns $x_i$ and $y_{ij}$ {\cite{Networks}\cite{Suppress}}
\begin{eqnarray}
  x_i&=&p_i\prod_{j=1}^{n}[r_{ji}y_{ji}g_j(x_j)-r_{ji}+1] \label{eqna1}\\
  y_{ij}&=&\frac{x_i}{r_{ji}y_{ji}g_j(x_j)-r_{ji}+1}, \label{eqna2}
\end{eqnarray}

where $r_{ji}\geq0$ is the fraction of {the} nodes in network $N_i$ that directly depends on nodes of
network $N_j$ and $r_{ii}=0$, {$x_i$ represents the fraction of the nodes that survive in network $N_i$ after removing all the nodes affected by the initial attack and the nodes depending on the failed nodes in other networks, $y_{ij}$ is the fraction of the survived nodes in network $N_i$ after the damage from all the networks connected to network $N_i$ except the network $N_j$.} We can analytically compute that the fraction $P_{\infty,k}=x_kg_k(x_k)$ of nodes in the giant component of network $N_k$ as a function of $p_i$ and $q_i$. Specially, if $n=2$, equation (\ref{eqna2}) {yields $y_{12}=p_1$, $y_{21}=p_2$}. Equations (\ref{eqna1}) can be simplified as:
\begin{equation}\label{eqna21}
  x_1=p_1(r_{21}p_2g_2(x_2)-r_{21}+1)
\end{equation}
\begin{equation}\label{eqna22}
  x_2=p_2(r_{12}p_1g_1(x_1)-r_{12}+1).
\end{equation}

If the process of cascade is a first order percolation transition, there is a single step discontinuity at the threshold $p_c$, which is also a function of $q_i$ \cite{Reduce}; otherwise, we let $p_c=0$. According to equation (\ref{eq:payoff_i}), we calculate the {expectation of network $N_i$'s payoff $Y_i$} by equation (\ref{eq:Yiint}).
\begin{widetext}
\begin{equation}
\label{s.long}
Y_i=\int_{p_c}^1\dots\int_{p_c}^1I_i(P_{\infty,i}(q_i,\dots,q_n,p_1,\dots,p_n))\prod_{j=1}^{n}\psi_j(p_j)\mathrm{d}p_1\dots\mathrm{d}p_n
-O_i(\langle k \rangle_i). \label{eq:Yiint}
\end{equation}
\end{widetext}

From (\ref{eq:payoff_system}), we can compute the social welfare $Y=\sum_{i=1}^{n}Y_i$ of the whole system. When each entity $i$ chooses strategy $q_i$, the payoff of entity $i$ is denoted by $Y_i(q)$ and the payoff of the system is denoted by $Y(q)$.

Before presenting our main results, we give the following definitions. A strategy profile $q^{*}\in Q$ achieves optimal social welfare if
\begin{equation}\label{social}
  \forall q\in Q; \quad Y(q^{*})\geq Y(q).
\end{equation}
A strategy profile $q^{*}\in Q$ is a Nash equilibrium if
\begin{equation}\label{social}
  \forall i,\quad q_i\in Q_i; \quad Y_i(q^{*}_i,q^{*}_{-i})\geq Y_i(q_i,q_{-i}).
\end{equation}

In our model, when the average degree is fixed, {we can get that} $O_i(\langle k \rangle_i)$ is unchanged about $q_i$, this is a positive-sum game. In equations (\ref{eqna1}) and (\ref{eqna2}), $g_i$ are {strict} increasing functions on $[0,1]$. We can prove that all $P_{\infty,i}$ are positively correlated in interdependent networks. Since all $O_i$ are fixed, according to {equation} (\ref{eq:Yiint}), $Y_i$ are positively correlated too. This indicates that it will not reduce other entities' payoffs when one entity changes his strategy profile to improve his own payoff. {It is easy to see, that at Nash equilibrium the system reaches the optimal social welfare.} Therefore, in such {a} scenario, the individual payoff of each entity is maximized at the same point as the optimal social welfare. The cooperation {among} different networks can be reached.

For example, for the game {among} interdependent scale-free (SF) networks whose strategy space is the power of degree distribution, a higher power leads to an improvement of the system's robustness \cite{catastrophic}. Due to the above conclusion, when the average degree is fixed, different networks can cooperate to improve the power of distribution as high as possible and reach the optimal social welfare.

However, if each $O_i$ is not fixed, the cooperation among different networks may be unattainable. In fact, {we will prove in the following that the cooperation is unreachable if the real variables $q_i$ can range in some interval continuously (that is, this is a continuous game).} {Without} loss of generality, we can assume that {the} payoff functions are differential. Then the necessary condition for a pure strategy Nash equilibrium is ${\partial Y_i}/{\partial q_i}=0$. The necessary condition for the optimal social welfare in this game is ${\partial Y}/{\partial q_i}=0$.

We proceed to analyze the game when the average degree can be adjusted by $q_i$ of each network $N_i$. When the payoff of this system achieves its Nash equilibrium, {i.e.,} $Y_i=Y^{\max}_i$, we have ${\partial Y_i}/{\partial q_i}=0$. According to equation (\ref{eq:payoff_i}), we have
\begin{equation}\label{partialYinash}
\frac{\partial Y_i}{\partial q_i}=\frac{\partial E(I_i(P_{\infty,i}))}{\partial q_i}-\frac{\partial O_i(\langle k \rangle_i)}{\partial q_i}=0.
\end{equation}
Note that $\langle k \rangle_i$ does not rely on $q_j$ if $j\neq i$, leading to $\partial O_i/\partial q_j=0$. At the Nash equilibrium, we {have}
\begin{eqnarray}
\nonumber\frac{\partial Y}{\partial q_j}&=&\sum_{i=1}^{n}\frac{\partial E(I_i(P_{\infty,i}))}{\partial q_j}-\sum_{i=1}^{n}\frac{\partial O_i(\langle k \rangle_i)}{\partial q_j}\\
&=&\sum_{i\neq j}\frac{\partial E(I_i(P_{\infty,i})))}{\partial q_j}.\label{partialYnash}
\end{eqnarray}

Since $\langle k \rangle_i$ is not fixed about $q_i$, ${\partial E(I_i(P_{\infty,i}))}/{\partial q_i}={\partial O_i(\langle k \rangle_i)}/{\partial q_i}\neq 0$. Due to the interdependency, the incomes of $n$ networks, which are decided by the robustness of the system, are positively correlated. ${\partial E(I_i(P_{\infty,i})))}/{\partial q_j}$ have the same sign and do not equal to $0$, leading to ${\partial Y}/{\partial q_j}\neq0$. Therefore, Nash equilibrium (since the concrete payoff function is {not} given here, the pure strategy Nash equilibrium may not exist, however, our model can be easily extended to the mixed strategy game in which Nash equilibrium always exists \cite{Gametheory}) deviates from {the} optimal social welfare. The cooperation among different interdependent networks can not be reached. The rationality of different entities {makes the system} inherently deficient.

\begin{table*}[t]
  \centering
  \begin{tabular}{|c|cccccc|}
  \hline
  \samepage
  \diagbox{$q_1$}{$(Y_1,Y_2)$}{$q_2$} & 5 & 6 & \underline{7} & 8 & \underline{9} & 10 \\
  \hline
  5 & (0.340,0.340) & (0.356,0.340) & (0.367,0.351) & (0.374,0.350) & (0.379,0.347) & (0.383,0.343) \\
  6 & (0.340,0.356) & (0.366,0.366) & (0.376,0.368) & (0.383,0.367) & (0.388,0.364) & (0.392,0.360) \\
  \underline{7} & (0.351,0.367) & (0.368,0.376) & \underline{(0.379,0.379)} & (0.386,0.378) & (0.391,0.375) & (0.395,0.371) \\
  8 & (0.350,0.374) & (0.367,0.383) & (0.378,0.386) & (0.385,0.385) & (0.390,0.382) & (0.394,0.378) \\
  \underline{9} & (0.347,0.379) & (0.364,0.388) & (0.375,0.391) & (0.382,0.390) & \underline{(0.387,0.387)} & (0.391,0.383) \\
  10 & (0.343,0.383) & (0.360,0.392) & (0.395,0.371) & (0.378,0.394) & (0.383,0.391) & (0.386,0.386) \\
  \hline
  \end{tabular}
  \caption{Game Matrix for two totally interdependent RR networks: the two components of the vectors in matrix are the payoffs of two entities with corresponding strategy profile {for $I_i(x)=x$ and $O_i(x)=0.008x$}. The Nash equilibrium and optimal social welfare are underlined in the matrix. {It can be seen that there is a notable gap between the Nash equilibrium $q_1=q_2=7$ and social optimum $q_1=q_2=9$.}}\label{RRtable}
\end{table*}

\begin{figure}[!]
  \centering
  \subfigure[]{
  \label{figurea}
  \includegraphics[scale=0.22]{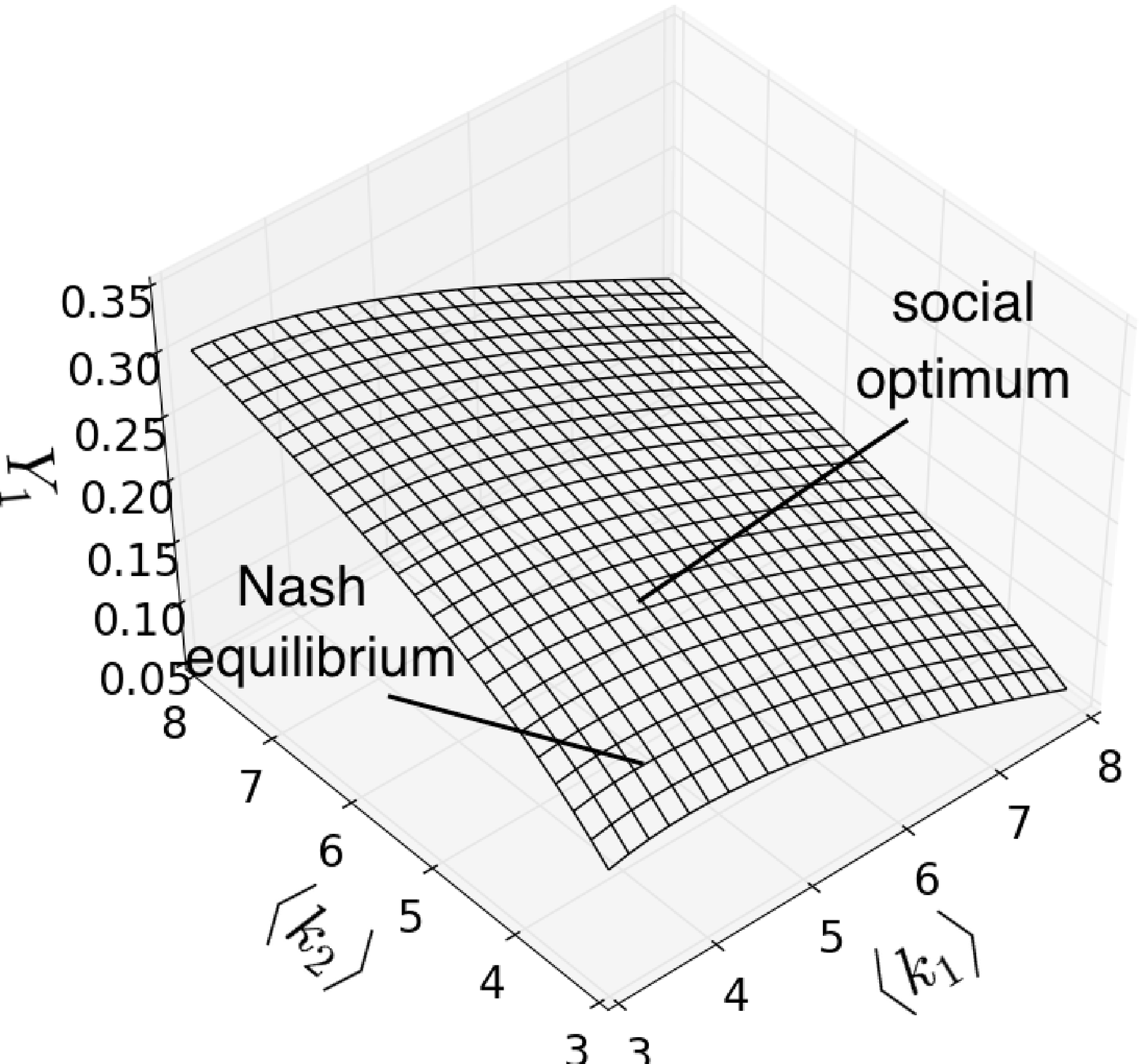}}
  \subfigure[]{
  \label{figureb}
  \includegraphics[scale=0.22]{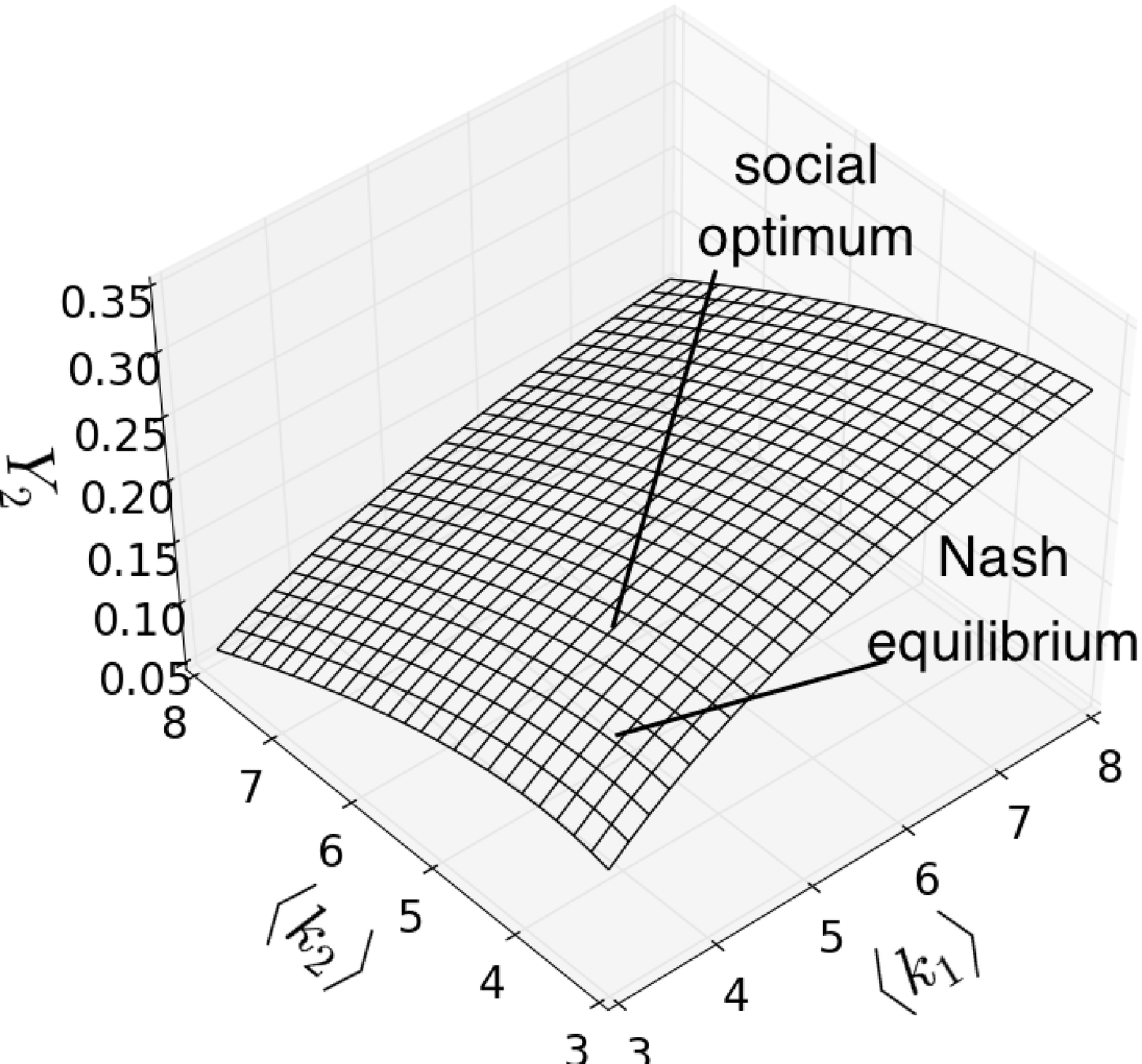}}
  \caption{{(Colour on-line)} For two totally coupled ER networks whose strategy space is the set of their average degrees $\{\langle k\rangle_1,\langle k\rangle_2 \}$, the payoff $Y_1$ {(figure (a))} and $Y_2$ {(figure (b))} are shown as functions of {$\langle k\rangle_1$ and $\langle k\rangle_2$ with $I_i(x)=x$ and $O_i(x)=0.05x$}. By numerical simulations, we get that this game reaches its social optimum at $\langle k\rangle_1=\langle k\rangle_2=4.37$ with $Y=0.54$ and reaches its Nash equilibrium at $\langle k\rangle_1=\langle k\rangle_2$=3.09 with $Y$=0.49. We can see that there is a {notable} gap between Nash equilibrium and social optimum.}\label{ERY}
\end{figure}

{If $Y_i=Y_i^{\max}$ is the payoff $Y_i$ of network $N_i$ at the Nash equilibrium and $Y=Y_{\max}$ at the social optimum, we define $\Delta$ as
\begin{equation}
\Delta=\frac{Y_1^{\max}+Y_2^{\max}}{Y_{\max}}\leq1
\end{equation}
to evaluate this game. The higher ${\Delta}$ is, the closer the payoffs {at} Nash equilibrium and social optimum are.}

Next we analyze our model for concrete interdependent networks {and strategy space}. {For the convenience of our numerical validation but without loss of generality, we set $I_i$ and $O_i$ to be linear functions in the following experiments.} For coupled Random Regular (RR) networks, we set $q_i\in{Z^{+}}$ to be their average degree $\langle k\rangle_i$ ($i\in\{1,2\}$). Then we have $G_i(x,q_i)=x^{q_{i}}$, $H_i=x^{q_{i}-1}$ {and $P_{\infty,i}=x_ig_i(x_i)$ where $g_i$ {are given} in equation (\ref{eq:gi}).} {Set $I_i(x)=x$ and $O_i(x)=0.008x$} in equation (\ref{eq:payoff_i}) and the distributions of $p_1$ and $p_2$ {to} be uniform distributions. Since the average degree of RR network {can only} be integer, the game {between two} interdependent RR networks is discrete. By numerical validation, we have the {payoff matrix} and {obtain} the Nash equilibrium and social optimum of this discrete game. TABLE.\ref{RRtable} is the game matrix for the case where two networks are totally interdependent, {i.e.,} $r_{12}=r_{21}=1$ {in equations (10) and (11)}. From this matrix, we can calculate that $q_1=q_2=9$ {are the strategy} in the social optimum and $q_1=q_2=7$ {are those in} the Nash equilibrium. Similarly, we can get the Nash equilibrium and social optimum for partially interdependent {RR} networks.

For two coupled ER networks, we set $q_i$ to be their average degree $\langle k\rangle_i$ ($i\in\{1,2\}$). Then we have $G_i(x,q_i)=H_i(x,q_i)=exp[q_i(x-1)]$. {Similarly to the first example, we set $I_i$ and $O_i$ to be linear functions whose coefficients are $1$ and $0.05$, {respectively}. We also set the {distributions} of $p_i$ to be uniform.} As FIG.\ref{ERY} shows, we can solve $Y_1$ and $Y_2$ as functions of $q_1$ and $q_2$. By numerical {simulations}, we can {obtain} the Nash equilibrium and social optimum for this game. For instance, in the case of two totally coupled ER networks, we get that this {system} reaches its social optimum at $\langle k\rangle_1=\langle k\rangle_2$=4.37 with $Y$=0.54 and reaches its Nash equilibrium at $\langle k\rangle_1=\langle k\rangle_2$=3.09 with $Y$=0.49.

\begin{figure}[h]
  \centering
  \subfigure[]{
  \label{rry}
  \includegraphics[scale=0.20]{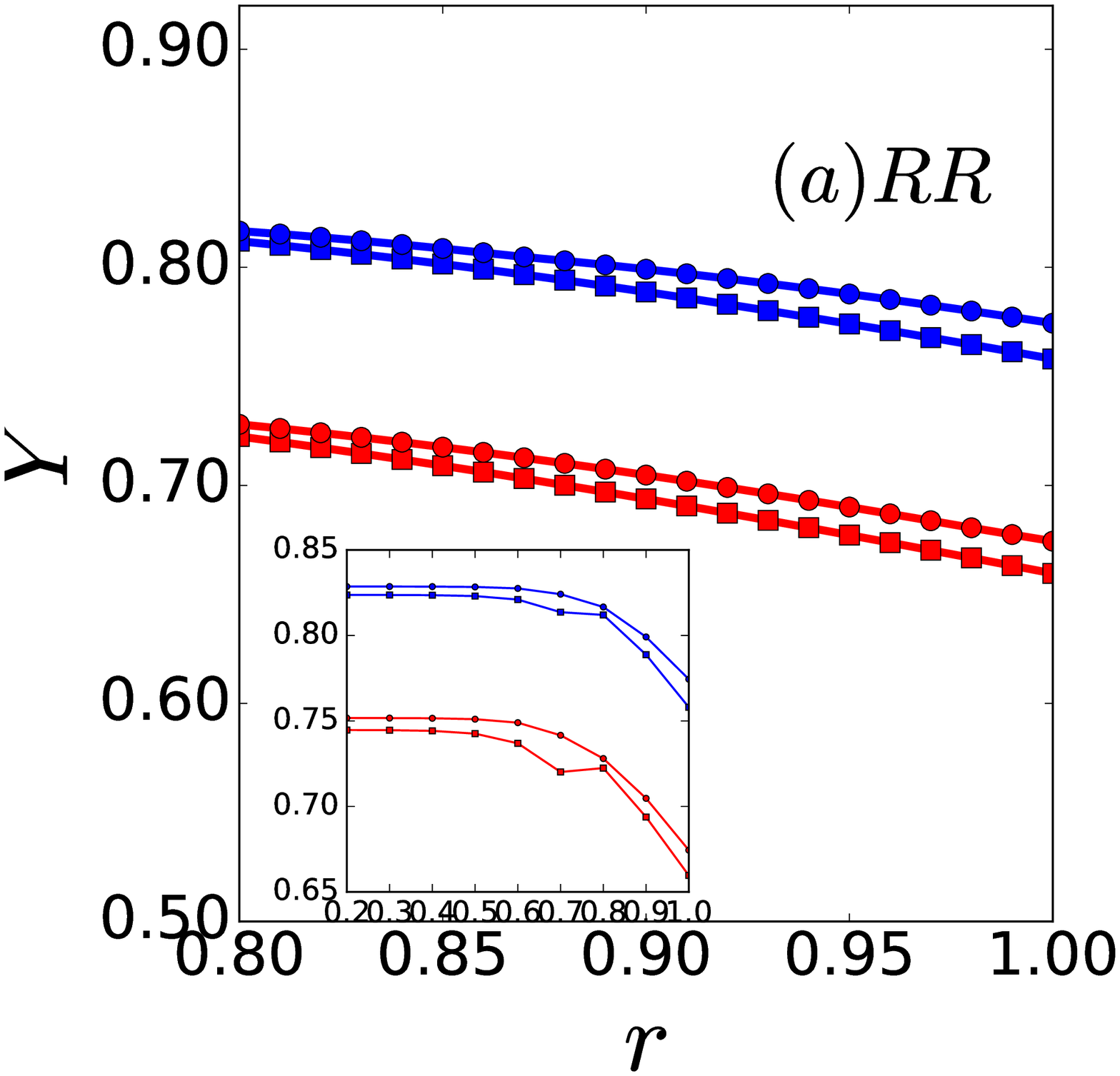}}
  \subfigure[]{
  \label{ery}
  \includegraphics[scale=0.20]{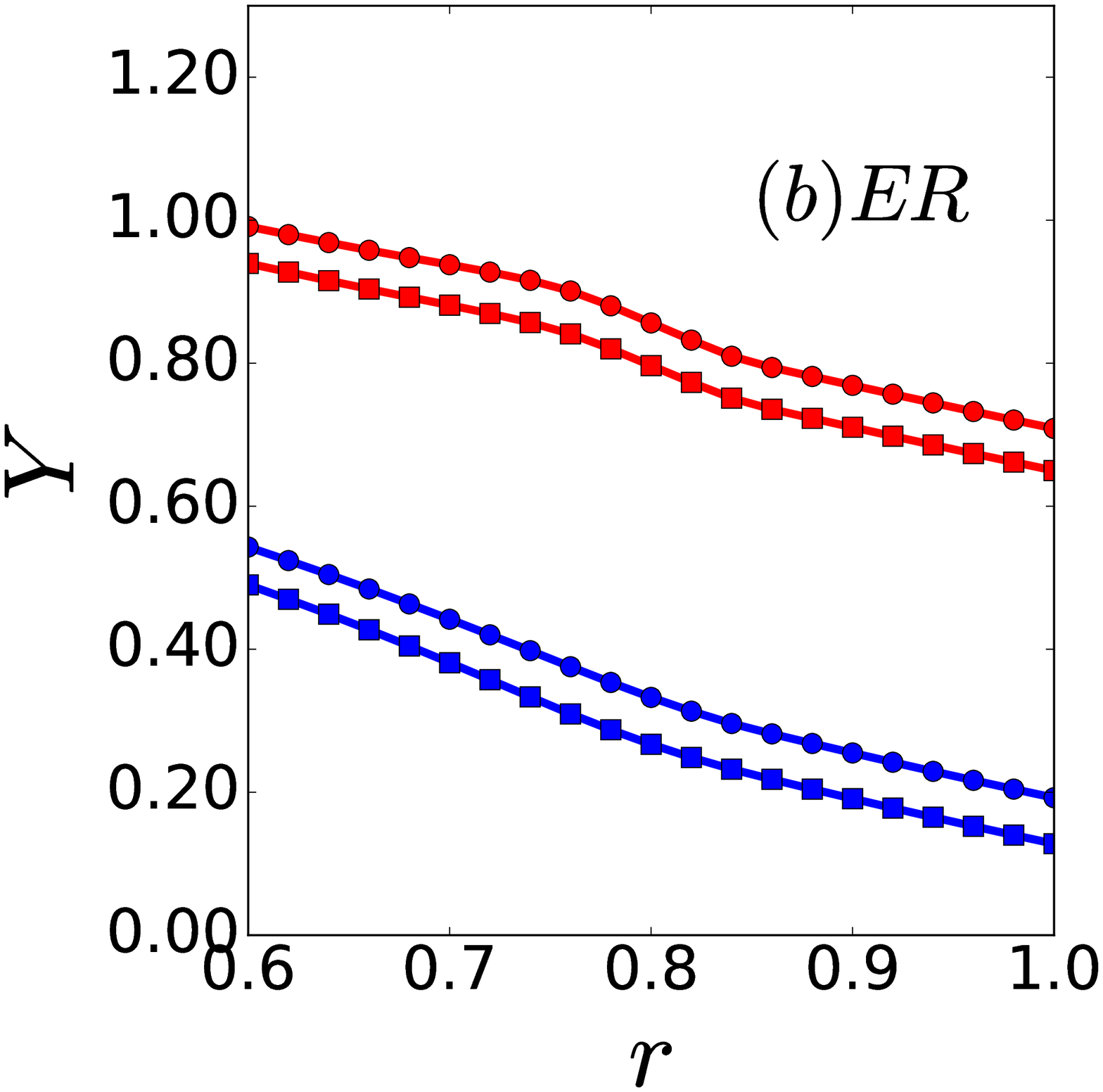}}
  \subfigure[]{
  \label{rrd}
  \includegraphics[scale=0.20]{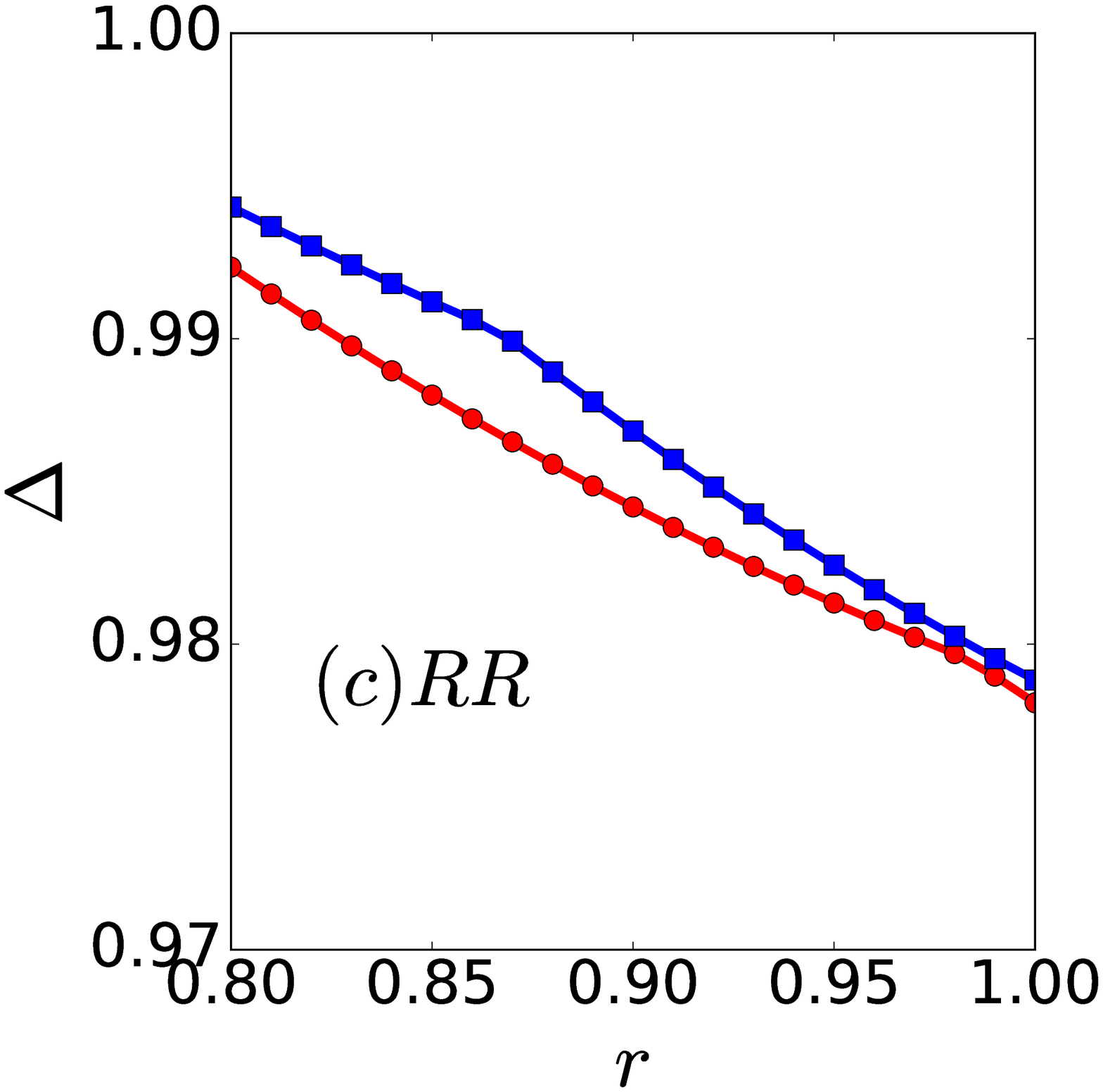}}
  \subfigure[]{
  \label{erd}
  \includegraphics[scale=0.20]{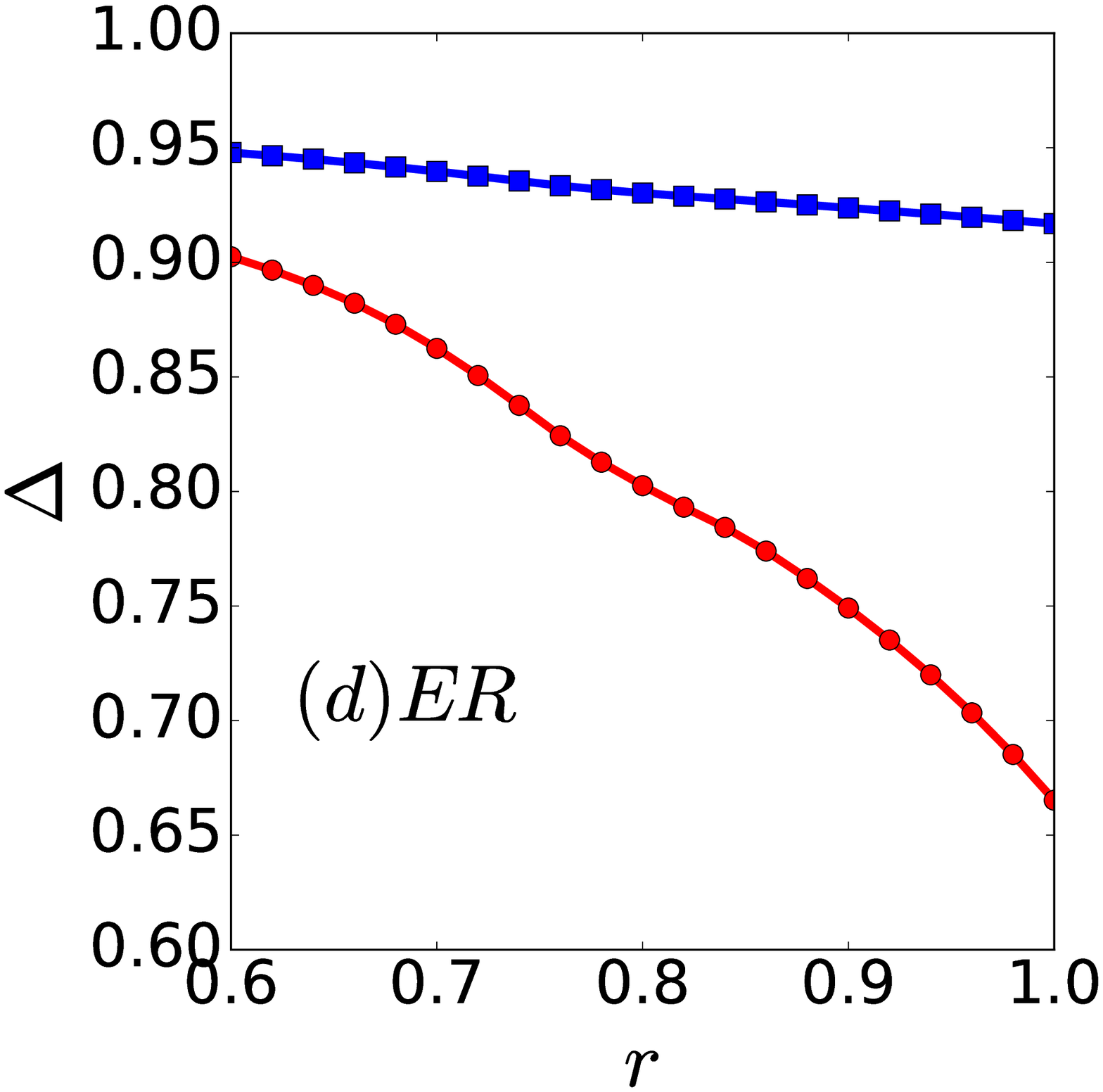}}
  \caption{{(Colour on-line)} For two coupled RR and ER {networks} whose strategy spaces are average degree, $Y_1^{\max}+Y_2^{\max}$, $Y_{\max}$ and $\Delta$ are shown as functions of coupled strength $r$ {with $I_i(x)=x$, $O_i(x)=cx$ and the distribution of $p_i(i\in\{1,2\})$ being uniform distribution. Figure (a) and figure (c) are $Y_{\max}-r$, $(Y_1^{\max}+Y_2^{\max})-r$ and $\Delta-r$ graphs for coupled RR networks with $c=0.014$ and $c=0.008$, respectively. {Figure (b) and figure (d) are $Y_{\max}-r$, $(Y_1^{\max}+Y_2^{\max})-r$ and $\Delta-r$ graphs for coupled ER networks with $c=0.05$ and $c=0.03$, respectively. Red square, red circle, blue square and blue circle curves correspond to $Y_1^{\max}+Y_2^{\max}$ at $c=0.014$, $Y_{\max}$ at $c=0.014$, $Y_1^{\max}+Y_2^{\max}$ at $c=0.008$, $Y_{\max}$ at $c=0.008$ in figure (a), and correspond to $Y_1^{\max}+Y_2^{\max}$ at $c=0.05$, $Y_{\max}$ at $c=0.05$, $Y_1^{\max}+Y_2^{\max}$ at $c=0.03$, $Y_{\max}$ at $c=0.03$ in figure (b). Red circle and blue square curves correspond to $\Delta$ at $c=0.014$ and $c=0.008$ in figure (c), and correspond to $\Delta$ at $c=0.05$ and $c=0.03$ in figure (d).} It can be seen from the subfigure of figure (a) that $Y_1^{\max}+Y_2^{\max}$ is not strictly decreasing in $r$.} We can see that reducing the coupled strength leads to {a remarkable} improvement of $Y_{\max}$, $Y_1^{\max}+Y_2^{\max}$ and $\Delta$.}\label{coupleds}
\end{figure}

Set $r_{12}=r_{21}=r\leq1$ in equations (\ref{eqna21}) and (\ref{eqna22}), we evaluate our model for interdependent networks with different coupled strength (i.e., with different $r$) and {different $I_i$ and $O_i$}. As FIG.\ref{coupleds} shows, we have that reducing the coupled strength of interdependent networks leads to an improvement of $Y_{\max}$, $Y_1^{\max}+Y_2^{\max}$ and $\Delta$. It is worth mentioning that, since the game {between two} interdependent RR networks is discrete, the Nash equilibrium, which only can be integer, does not change continuously about $r$ and has discontinuity. Thus the $Y_1^{\max}+Y_2^{\max}$ and $\Delta$ are not strictly decreasing about $r$ {(as the subfigure of FIG.\ref{rry} shows)}. However, in general, the system is more profitable and efficient as a result of reducing the coupled strength.

\begin{figure}[!]
  \centering
  \subfigure[]{
  \label{Prr}
  \includegraphics[scale=0.2]{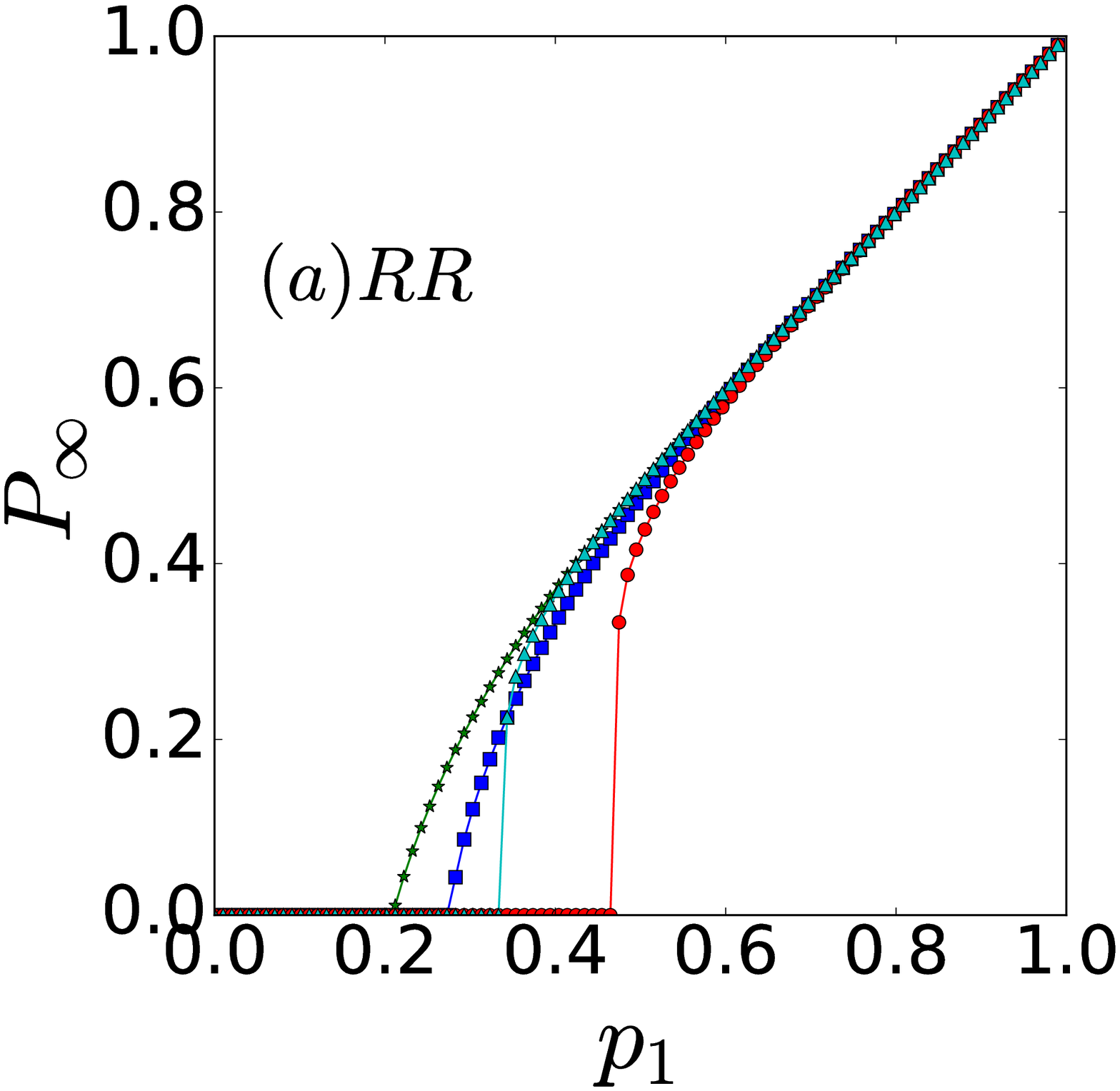}}
  \subfigure[]{
  \label{Per}
  \includegraphics[scale=0.2]{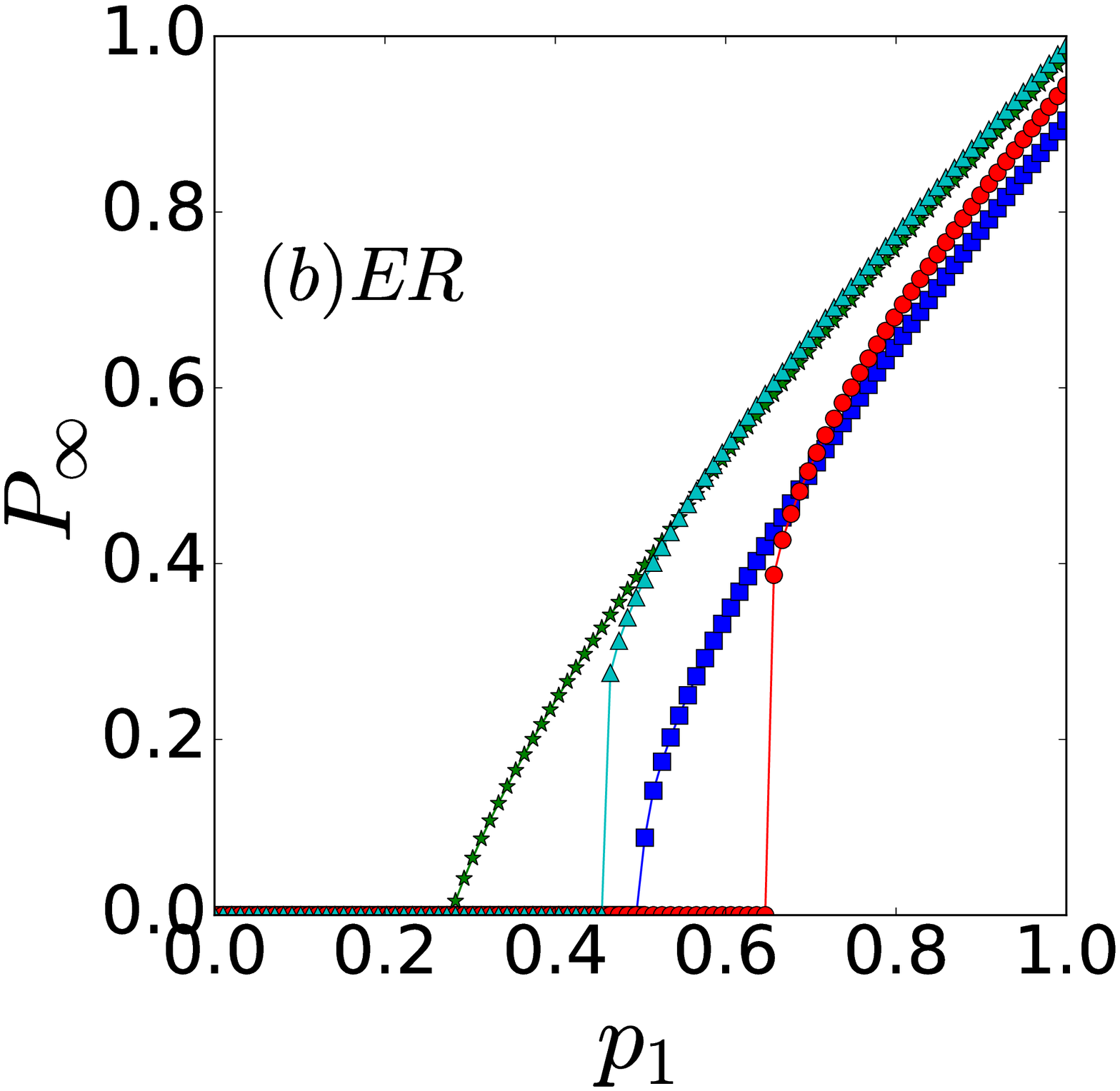}}
  \caption{{(Colour on-line) $P_\infty$-$p$ graph, of two coupled RR networks and ER networks at Nash equilibrium and social optimum with $I_i=x$, $O_i=0.008x$ for figure (a), $I_i=x$, $O_i=0.05x$ for figure (b) and $r=0.6$, $r=1$ respectively.} {$P_\infty$}, the fraction of nodes in the giant component of network $N_1$, is shown as a function of {the fraction $p_1$ of remaining nodes in $N_1$. Squire, star, circle and triangle curves correspond to Nash equilibrium at $r=0.6$, social optimum at $r=0.6$, Nash equilibrium at $r=1$ and social optimum at $r=1$, respectively.} We can see that {the threshold of percolation transition is higher and $P_{\infty}$ is lower at Nash equilibrium in the same scenario, and that} the system is more vulnerable at Nash equilibrium than those at social optimum.}\label{Pp}
\end{figure}

\begin{table}[]
 \centering
 \begin{tabular}{|c|c|c|c|c|c|}
  \hline
  RR & star-like & chain-like & ER & star-like & chain-like\\
  \hline
  $\Delta$ & 1.0 & 0.89 & $\Delta$ & 0.56 & 0.52\\
  \hline
 \end{tabular}
 \caption{$\Delta$ is shown for {five} interdependent RR networks and ER networks with different dependency topology. We can see that the $\Delta$ is higher for star-like system {comparing to chain-like system.}}\label{topology}
\end{table}

\begin{figure}[h]
  \centering
  \subfigure[Chain-like system]{
  \includegraphics[scale=0.15]{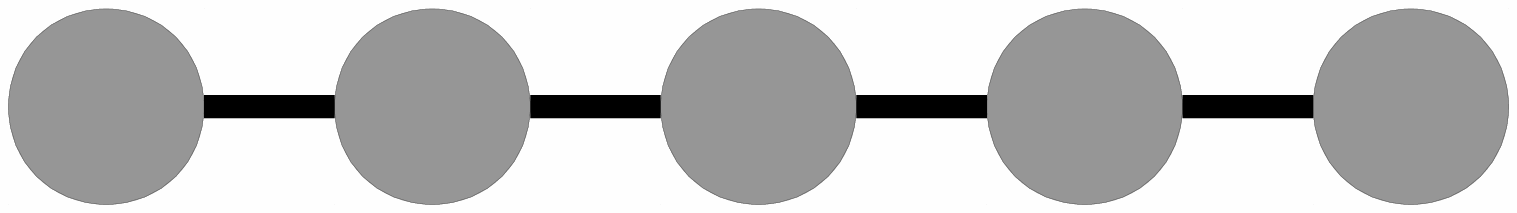}}
  \subfigure[Star-like system]{
  \includegraphics[scale=0.15]{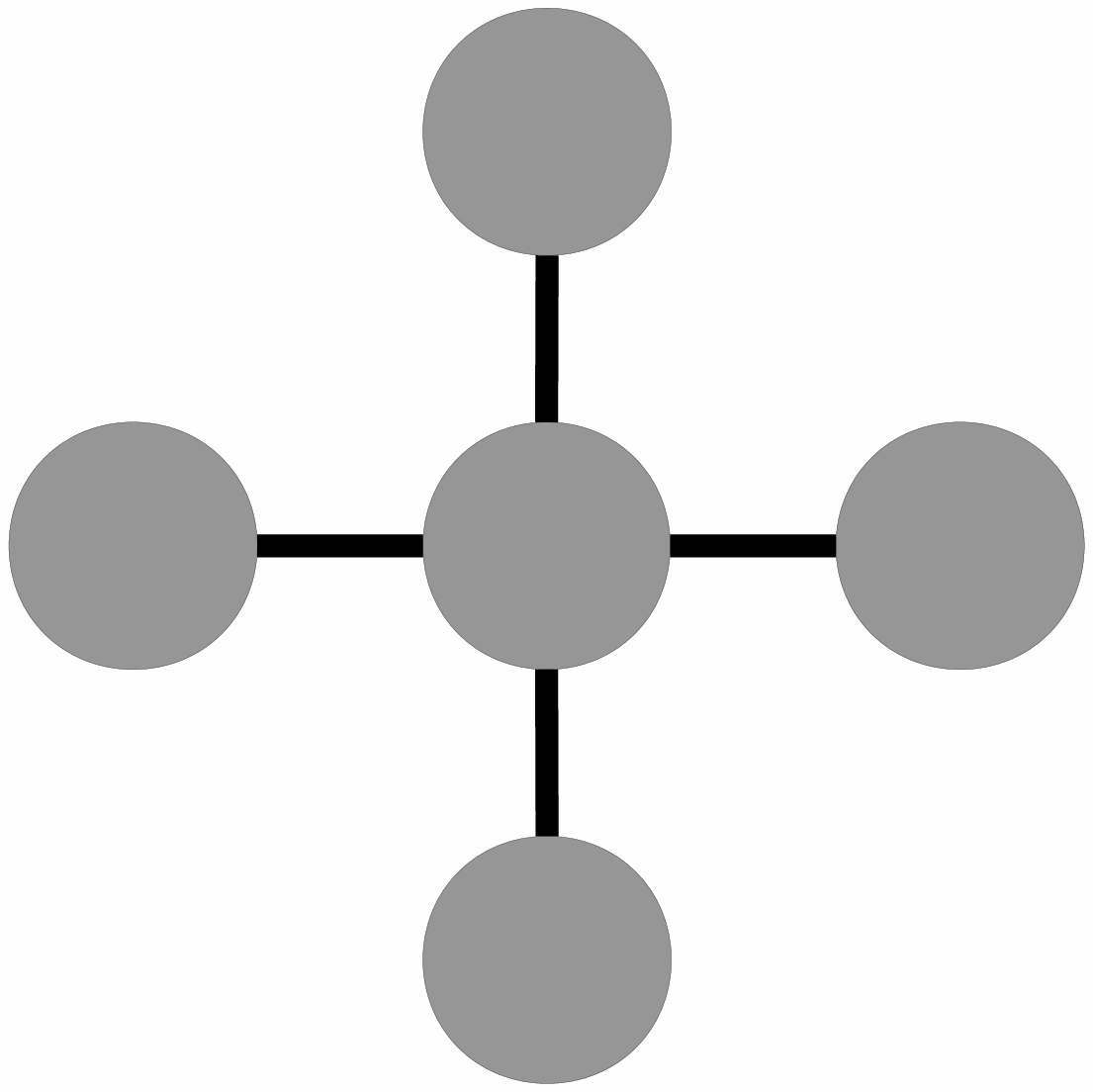}}
  \caption{Two types of {essential} systems composed of five coupled networks}\label{dependent}
\end{figure}

From the above examples of RR networks and ER networks, we can see that the rational behaviors will make the system {away from} the optimal robustness and {render it} more vulnerable {in some cases}. As FIG.\ref{Pp} shows, we get the $P_{\infty}-p$ graph at {the} Nash equilibrium and social optimum for $r=1$ and $r=0.6$, {respectively}. The threshold is higher and $P_{\infty}$ is lower at Nash equilibrium in the same scenario. Note that the system of interdependent networks is more vulnerable at Nash equilibrium as a result of rationality.

It is important to find a way of reducing gap between optimal social welfare and Nash equilibrium for the game {among} interdependent networks. Besides reducing coupled strength, surprisingly, we find that the gap is narrower for the system with suitable dependency topology. {Here we} test our model for {two systems composed of five} interdependent networks with different dependency topology. Let all coupled networks be totally interdependent. According to {the} one-to-one correspondence, randomly removing $1-p_i$ fractions of nodes from each network $N_i$ {is} equivalent to a single attack on one of the networks which removes $1-p=1-\prod_{i=1}^{5}p_i$ fraction of nodes. Set the distribution of $p$ to be uniform and $I_i$ and $O_i$ be linear functions with coefficients $1$ and $3.5\times10^{-4}$ in equations (\ref{eq:payoff_i}) for RR networks and {coefficients} $1$ and $2.5\times10^{-4}$ for ER networks. We validate the cases of chain-like and star-like system formed from five networks (See FIG.\ref{dependent}). By numerical simulations, we {obtain} that the star-like {oligarchic} dependency topology tends to reduce the gap (See TABLE.\ref{topology}).

Summarizing, in this paper, we study the influence of rational behavior on {system robustness}. We reveal that, in the continuous game, the cooperation among different {interdependent} networks is reachable only when the average degree of each network is fixed in strategy space. While in general, there is a huge gap between the Nash equilibrium and optimal social welfare as a result of rationality which makes {the system} inherent deficient. We reveal and validate some factors (including weakening coupled strength of interdependent networks and designing suitable topology dependency of system) that help reduce the gap and deficiency.

Interdependent networks exist in all aspects of our life, nature and technology. The game {among} interdependent networks is more complicated in real world since besides the fraction of giant component and average degree, the {utility function} may rely on many factors and need further investigation. It is of great importance to find other efficient ways of reducing {the system's vulnerability and the gap between Nash equilibrium and optimal social welfare}.

\acknowledgments
This work is supported by NSFC project {under grant} 61528105 {and Zhejiang Provincial Natural Science Foundation of China under grant LR16F020001.} We thank Prof. Yang-Yu Liu at Harvard university for his valuable suggestion, Zidong Yang, Yongtao Zhang and {Hanyuan Liu} at Zhejiang university for their insightful discussions.

\bibliographystyle{unsrt}%

\begin{thebibliography}{0}
{
\bibitem{catastrophic}
  \Name{Buldyrev S. V. \and Parshani R. \and Paul G. \and Stanley H. E. \and Havlin S.}
  \Book{Nature}
  \Vol{464}
  \Page{1025--1028}
  \Year{2010}
  \Publ{Nature Publishing Group}.

\bibitem{Networks}
  \Name{Gao J. \and Buldyrev S. V. \and Stanley H. E. \and Havlin S.}
  \Book{Nature physics}
  \Vol{8}
  \Page{40--48}
  \Year{2012}
  \Publ{Nature Publishing Group}.

\bibitem{kenett2015networks}
  \Name{Kenett D. Y. \and Perc M. \and Boccaletti S.}
  \Book{Chaos, Solitons \& Fractals}
  \Vol{80}
  \Page{1--6}
  \Year{2015}
  \Publ{Elsevier}.

\bibitem{Towards}
  \Name{Schneider C. M. \and Yazdani N. \and Ara{\'u}jo N. A. \and Havlin S. \and Herrmann H. J.}
  \Book{Scientific reports}
  \Vol{3}
  \Year{2013}
  \Publ{Nature Publishing Group}.

\bibitem{Reduce}
  \Name{Parshani R. \and Buldyrev S. V. \and Havlin S.}
  \Book{Physical review letters}
  \Vol{105}
  \Page{048701}
  \Year{2010}
  \Publ{APS}.

\bibitem{Multisupport}
  \Name{Shao J. \and Buldyrev S. V. \and Havlin S. \and Stanley H. E.}
  \Book{Physical Review E}
  \Vol{83}
  \Page{036116}
  \Year{2011}
  \Publ{APS}.

\bibitem{Robustness}
  \Name{Gao J. \and Buldyrev S. V. \and Havlin S. \and Stanley H. E.}
  \Book{Physical Review Letters}
  \Vol{107}
  \Page{195701}
  \Year{2011}
  \Publ{APS}.

\bibitem{Recovery}
  \Name{Di Muro M. A. \and La Rocca C. E. \and Stanley H. E. \and Havlin S. \and Braunstein L. A.}
  \Book{Scientific reports}
  \Vol{6}
  \Year{2016}
  \Publ{Nature Publishing Group}.

\bibitem{Recentadvances}
  \Name{Shekhtman L. M. \and Danziger M. M. \and Havlin S.}
  \Book{Chaos, Solitons \& Fractals}
  \Vol{90}
  \Page{28--36}
  \Year{2016}
  \Publ{Elsevier}.

\bibitem{wang2012evolution}
  \Name{Wang Z. \and Szolnoki A. \and Perc M.}
  \Book{Europhysics Letters}
  \Vol{97}
  \Page{48001}
  \Year{2012}
  \Publ{IOP Publishing}.

\bibitem{wang2013interdependent}
  \Name{Wang Z. \and Szolnoki A. \and Perc M.}
  \Book{Scientific Reports}
  \Vol{3}
  \Page{1183}
  \Year{2013}.

\bibitem{wang2015evolutionary}
  \Name{Wang Z. \and Wang L. \and Szolnoki A. \and Perc M.}
  \Book{The European Physical Journal B}
  \Vol{88}
  \Page{1--15}
  \Year{2015}
  \Publ{Springer}.

\bibitem{wang2014self}
  \Name{Wang Z. \and Szolnoki A. \and Perc M.}
  \Book{New Journal of Physics}
  \Vol{16}
  \Page{033041}
  \Year{2014}
  \Publ{IOP Publishing}.

\bibitem{wang2013optimal}
  \Name{Wang Z. \and Szolnoki A. \and Perc M.}
  \Book{Scientific reports}
  \Vol{3}
  \Year{2013}
  \Publ{Nature Publishing Group}.

\bibitem{jiang2013spreading}
  \Name{Jiang L. L. \and Perc M.}
  \Book{Scientific reports}
  \Vol{3}
  \Year{2013}
  \Publ{Nature Publishing Group}.

\bibitem{szolnoki2013information}
  \Name{Szolnoki A. \and Perc M.}
  \Book{New Journal of Physics}
  \Vol{15}
  \Page{053010}
  \Year{2013}
  \Publ{IOP Publishing}.

\bibitem{nowak1992evolutionary}
  \Name{Nowak M. A. \and May R. M.}
  \Book{Nature}
  \Vol{359}
  \Page{826--829}
  \Year{1992}.

\bibitem{hauert2004spatial}
  \Name{Hauert C. \and Doebeli M.}
  \Book{Nature}
  \Vol{428}
  \Page{643--646}
  \Year{2004}
  \Publ{Nature Publishing Group}.

\bibitem{Economic}
  \Name{Schweitzer F. \and Fagiolo G. \and Sornette D. \and Vega-Redondo F. \and Vespignani A. \and White D. R.}
  \Book{science}
  \Vol{325}
  \Page{422--425}
  \Year{2009}
  \Publ{American Association for the Advancement of Science}.

\bibitem{Modelling}
  \Name{Rosato V. \and Issacharoff L. \and Tiriticco F. \and Meloni S. \and Porcellinis S. \and Setola R.}
  \Book{International Journal of Critical Infrastructures}
  \Vol{4}
  \Page{63--79}
  \Year{2008}
  \Publ{Inderscience Publishers}.

\bibitem{Identifying}
  \Name{Rinaldi S. M. \and Peerenboom J. P. \and Kelly T. K.}
  \Book{IEEE Control Systems}
  \Vol{21}
  \Page{11--25}
  \Year{2001}
  \Publ{IEEE}.

\bibitem{Percolation}
  \Name{Callaway D. S. \and Newman M. E. \and Strogatz S. H. \and Watts D. J.}
  \Book{Physical review letters}
  \Vol{85}
  \Page{5468}
  \Year{2000}
  \Publ{APS}.

\bibitem{Modellinginterdependent}
  \Name{Rosato V. \and Issacharoff L. \and Tiriticco F. \and Meloni S. \and Porcellinis S. \and Setola R.}
  \Book{International Journal of Critical Infrastructures}
  \Vol{4}
  \Page{63--79}
  \Year{2008}
  \Publ{Inderscience Publishers}.

\bibitem{Random}
  \Name{Bollob{\'a}s B.}
  \Book{Modern Graph Theory}
  \Page{215--252}
  \Year{1998}
  \Publ{Springer}.

\bibitem{onecorrespondence}
  \Name{Gao J. \and Buldyrev S. V. \and Havlin S. \and Stanley H. E.}
  \Book{Physical Review E}
  \Vol{85}
  \Page{066134}
  \Year{2012}
  \Publ{APS}.

\bibitem{Gametheory}
  \Name{Fudenberg D. \and Tirole J.}
  \Book{Game theory}
  \Vol{393}
  \Year{1991}.

\bibitem{Aprimer}
  \Name{Gibbons R.}
  \Year{1992}
  \Publ{Harvester Wheatsheaf}.

\bibitem{Fractal}
  \Name{Shao J. \and Buldyrev S. V. \and Cohen R. \and Kitsak M. \and Havlin S. \and Stanley H. E.}
  \Book{Europhysics Letters}
  \Vol{84}
  \Page{48004}
  \Year{2008}
  \Publ{IOP Publishing}.

\bibitem{Spread}
  \Name{Newman M. E.}
  \Book{Physical review E}
  \Vol{66}
  \Page{016128}
  \Year{2002}
  \Publ{APS}.

\bibitem{Suppress}
  \Name{Brummitt C. D. \and D'Souza R. M. \and Leicht E. A.}
  \Book{Proceedings of the National Academy of Sciences}
  \Vol{109}
  \Page{E680--E689}
  \Year{2012}
  \Publ{National Acad Sciences}.
}
\end{thebibliography}

\end{document}